\documentclass[preprint]{imsart}
\RequirePackage[OT1]{fontenc}
\RequirePackage{amsthm,amsmath,mathrsfs,amssymb,amsfonts}
\RequirePackage[numbers]{natbib}
\RequirePackage[colorlinks,citecolor=blue,urlcolor=blue]{hyperref}
\usepackage{fullpage}
\usepackage{graphicx}

\startlocaldefs
\numberwithin{equation}{section}
\theoremstyle{plain}

\DeclareMathOperator*{\Discrete}{Discrete}
\DeclareMathOperator*{\Dirichlet}{Dirichlet}
\DeclareMathOperator*{\DP}{DP}
\DeclareMathOperator*{\Bet}{Beta}
\DeclareMathOperator*{\Uniform}{Uniform}
\DeclareMathOperator*{\Bernoulli}{Bernoulli}

\def\Prob{\mathbb{P}}
\def\xil{x_{il}}
\def\qik{q_{ik}}
\def\qij{q_{ij}}
\def\Qi{Q_i}
\def\thetakla{\theta_{kla}}
\def\thetakl{\theta_{kl}}
\def\thetak{\theta_k}
\def\dl{d_l}
\def\zil{z_{il}}
\def\zin{z_{i,l+1}}
\def\zione{z_{i1}}
\def\sil{s_{il}}
\def\sin{s_{i,l+1}}
\def\Qz{Q_0}
\def\qzk{q_{0k}}
\def\qzj{q_{0j}}
\def\xil{x_{il}}
\def\qik{q_{ik}}
\def\qij{q_{ij}}
\def\Qi{Q_i}
\def\thetakla{\theta_{kla}}
\def\thetakl{\theta_{kl}}
\def\thetak{\theta_k}
\def\dl{d_l}
\def\zil{z_{il}}
\def\zin{z_{i,l+1}}
\def\zione{z_{i1}}
\def\sil{s_{il}}
\def\sin{s_{i,l+1}}
\def\Qz{Q_0}
\def\qzk{q_{0k}}
\def\qzj{q_{0j}}
\def\xil{x_{il}}
\def\qik{q_{ik}}
\def\qij{q_{ij}}
\def\Qi{Q_i}
\def\thetakla{\theta_{kla}}
\def\thetakl{\theta_{kl}}
\def\thetak{\theta_k}
\def\dl{d_l}
\def\zil{z_{il}}
\def\zin{z_{i,l+1}}
\def\zione{z_{i1}}
\def\sil{s_{il}}
\def\sin{s_{i,l+1}}
\def\Qz{Q_0}
\def\qzk{q_{0k}}
\def\qzj{q_{0j}}

\endlocaldefs

\begin{document}

\begin{frontmatter}
\title{Modeling population structure under hierarchical Dirichlet processes}
\runtitle{Modeling population structure}

\begin{aug}
\author{\fnms{M.} \snm{De Iorio}\ead[label=e1]{m.deiorio@ucl.ac.uk}},
\author{\fnms{L.T.} \snm{Elliott}\ead[label=e3]{elliott@stats.ox.ac.uk}},
\author{\fnms{S.} \snm{Favaro}\thanksref{t1}\ead[label=e2]{stefano.favaro@unito.it}},
\author{\fnms{K.} \snm{Adhikari}\ead[label=e5]{k.adhikari@ucl.ac.uk}}
\and
\author{\fnms{Y.W.} \snm{Teh}\ead[label=e4]{y.w.teh@stats.ox.ac.uk}
\ead[label=u1,url]{http://www.foo.com}}

\runauthor{M. De Iorio, L. Elliott, S. Favaro, K. Adhikari and Y.W. Teh}

\thankstext{t1}{Also affiliated to Collegio Carlo Alberto, Moncalieri, Italy.}

\affiliation{University College London, University of Oxford, University of Torino, University College London and University of Oxford}

\address{Department of Statistical Science \\
Gower Street\\
London WC1E 6BT\\
United Kingdom\\
\printead{e1}}

\address{Department of Economics and Statistics\\
Corso Unione Sovietica 218/bis\\
10134 Torino\\
Italy\\
\printead{e2}}

\address{Department of Genetics, Evolution and Environment\\
Gower Street\\
London WC1E 6BT\\
United Kingdom\\
\printead{e5}}

\address{Department of Statistics\\
1 South Parks Road\\
Oxford OX1 3TG\\
United Kingdom\\
\printead{e3}
\printead{e4}}
\end{aug}

\begin{center}
\begin{abstract}
We propose a Bayesian nonparametric model to infer population admixture, extending the Hierarchical Dirichlet Process (HDP, Teh {\em et al.} 2006) to allow for correlation between loci due to Linkage Disequilibrium. Given multilocus genotype data from a sample of individuals, the model allows inferring 
%the demographic origin of an individual,
classifying individuals as unadmixed or admixed, inferring the number of subpopulations ancestral to an admixed population and the population of origin of chromosomal regions. Our model does not assume any specific mutation process and can be applied to most of the commonly used genetic markers. We present a MCMC algorithm to perform posterior inference from the model and discuss methods to summarise the MCMC output for the analysis of population admixture.
We demonstrate the performance of the proposed model in simulations and in a real application, using genetic data from the EDAR gene,
which is considered to be ancestry-informative due to well-known variations in allele frequency as well as phenotypic effects across ancestry. The structure analysis of this dataset leads to the identification of a rare haplotype in Europeans.
\end{abstract}
\end{center}



\end{frontmatter}

%%%%%%%%%%%%%%%%%%%%%%%%%%%%%%%%
%%%%%%%%%%%%%%%%%%%%%%%%%%%%%%%%
%%%%%%%%%%%%%%%%%%%%%%%%%%%%%%%%
%%%%%%%%%%%%%%%%%%%%%%%%%%%%%%%%

\section{Introduction}
Population stratification, or population structure, refers to the presence of a systematic difference in genetic markers' allele frequencies between subpopulations due to variation in ancestry. This phenomenon  arises from the bio-geographical distribution of human populations. The analysis of population structure  presents an important problem in population genetics and arises in many contexts.  It is central to the understanding of human migratory history and the genesis of modern populations \cite{Ros(02),Rei(09)}. The associated admixture analysis of individuals is important in correcting the confounding effects of population ancestry on gene mapping \cite{Zhu(08)} and association studies \cite{Price(10)}.  It is also useful in the analysis of gene flow in hybridization zones \cite{Fie(11)} and invasive species \cite{Ray(14)}, conservation genetics \cite{Was(07)} and domestication events \cite{Park(04)}. The establishment of inexpensive single nucleotide polymorphism (SNP) genotyping platforms
in recent years has facilitated collection of markers to assess genetic ancestry in human populations and in general to investigate genetic relationships in living organisms. This paper focuses on a particular form of population structure: admixture. Genetic admixtures occur when two or more previously isolated populations begin interbreeding, resulting in the introduction of new genetic lineages  into a population (e.g.\ the  African-American population).

Broadly speaking, the aim of population structure analysis based on genetic data are: (i) detection of population structure; (ii) estimating the number of subpopulations in a sample; (iii) assigning individual in subpopulations; (iv) defining the number of ancestral populations in admixed populations; (v) inferring ancestral population proportions to admixed individuals; (vi) identifying the genetic ancestry of distinct chromosomal segments within an individual. A variety of modelling approaches have been proposed in the literature. Two of the most widely used approaches are Principal Component Analysis (PCA) and model-based estimation of ancestry, mainly involving clustering techniques or hidden Markov models. The PCA approach has been used to infer population structure for several decades. In the PCA approach, the individuals' genotypes are projected onto a lower dimensional space so that the locations of individuals in the projected space reflects their genetic similarities \cite{Pat(06),Nov(08)}. It should be noted that the top principal components do not always capture population  structure but may reflect family relatedness, long range linkage disequilibrium or simply genotyping artefacts. Model based methods aim to reconstruct historical events and therefore to infer explicitly genetic ancestry (e.g. \cite{Pri(00),Tan(05),Ale(09)}).  In the structured association approach samples are assigned to sub-population clusters, possibly allowing for fractional membership.

An influential early approach is STRUCTURE by Pritchard et al.\ (2000)~\cite{Pri(00)}, which assumes that individuals come from one of $K$ sub-populations. Based on Bayesian mixture models, population membership and population specific allele frequencies are jointly estimated from the data. This simple framework can be extended to genetic admixtures, allowing individuals to have ancestry from more than one population. For each individual, STRUCTURE determines what proportion of the individuals' genome comes from each population, while the alleles at different loci are modelled conditionally independently given these  admixture proportions. Taking a Bayesian approach to inference, independent priors on the allelic profile parameters of each population are specified and posterior inference is performed through Markov chain Monte Carlo (MCMC). Various extensions to STRUCTURE have been proposed to address a number of shortcomings. 

One problem with STRUCTURE, which we address in this paper , is that  of admixture linkage disequilibrium among neighbouring loci. When individuals from different groups admix, their offspring€s' DNA become a mixture of the DNA from each admixing group. 
Chunks of  DNA are passed along through subsequent generations, up to the present day. Therefore, the genomes of the descendants contain segments of DNA inherited from each of the original populations. The shorter the distance between two loci, the higher  the
probability that the population of ancestry will be the same at these two loci.  This means that ancestry states are autocorrelated. The lengths of uninterrupted DNA segments inherited from each sub-population reflect how long ago the admixture event occurred. In general long uninterrupted
segments from each population imply a recent admixture event. The original version of  STRUCTURE did not deal with admixture linkage disequilibrium and as a result it is necessary to thin out tightly-linked loci to reduce correlations which can affect the quality of inference. In \cite{Fal(03)}, Falush et al.\ (2003) improved on this issue by introducing a module to model linkage locally among neighbouring loci, using a Markov model which segments each chromosome into contiguous regions with shared genetic ancestry.  This allows  \emph{local genetic ancestry} from genotype data to be inferred, as opposed to the \emph{global} admixture proportions in \cite{Pri(00)}.  Such local ancestry estimation gives more fine-grained information about the admixture process. In \cite{Pri(00)} and \cite{Fal(03)}, each population is characterised by an allele frequency profile and they assume that the alleles at each locus are independent.  Further dependence can be modelled by introducing dependence among allelic profiles due to common ancestry \cite{Fal(03)} or by using a Markov model for the alleles conditionally on the ancestral segmentation \cite{Tan(05)}.  Another class of models uses haplotypic profiles as richer representations for the allelic dependence within populations \cite{Har(13)}.

Another important statistical concern in admixture modelling, which we address in this paper, is  the determination of the number of ancestral populations.  In Pritchard et al.\ (2000)~\cite{Pri(00)}, this is achieved using a model selection criteria based on  MCMC estimates of the log marginal probabilities of the data and the Bayesian deviance information criterion, though it has been noted by \cite{Fal(03)} that such estimates are highly sensitive to prior specifications regarding the relatedness of populations.   See also Corander et al.\ (2003)~\cite{Cor(03)} and Evanno et al.\ (2005)~\cite{Eva(05)} for other parametric approaches to determining the number of populations.
One way in which such model selection can be sidestepped is by using Bayesian nonparametric models \cite{bnp}, which offer a flexible framework and do not require the specification of a fixed and finite model size (which, in our context, is the number of populations).  Rather, one assumes an unbounded potential model size, of which only a finite part is observed on a given finite dataset.  Such ideas have been applied to population structure inference by Huelsenbeck and Andolfatto \cite{Hue(07)} who used the Dirichlet Process (DP) to define a Bayesian nonparametric counterpart to the ``no-admixture" model of Pritchard et al.\ (2000)~\cite{Pri(00)} (see also  Dawson and Belkhir \cite{Daw(01)} and Pella and Masuda \cite{Pel(06)} for extensions to polyploid data), and by Sohn et al.\ (2012)~\cite{Soh(12)} who used infinite hidden Markov models \cite{Bea(02)} for modelling linkage disequilibrium in admixture modelling.

In this paper, we propose a method for  modelling population structure that simultaneously gives estimates of local ancestries, bypasses difficult model selection issues using Bayesian nonparametrics, and is designed to be computationally more scalable than current Bayesian nonparametric methods. Our approach is based on using the hierarchical Dirichlet process (HDP) \cite{Teh(06)} to non-parametrically model the unknown and uncertain number of populations without having to perform costly model selection.  Unlike \cite{Soh(12)}, we use a simplified transition model in which, during a transition event, the founder identities on either side of the transition are independent.  This transition model requires a linear rather than quadratic (in the number of founders) number of parameters, as well as a forward-backward algorithm which scales linearly in the number of extant populations while introducing less auxiliary variables which can slow down convergence.  

In Section~\ref{sec:method} we introduce our Bayesian nonparametric model, as well as a novel Markov chain Monte Carlo method which allows efficient exact inference using a retrospective slice sampling truncation scheme.  Section~\ref{sec:simulation} describes results of simulation studies, and Section~\ref{sec:experiments} describes population structure analyses of genotype data from the EDAR gene region.  Section~\ref{sec:discussion} closes with a discussion of our findings as well as potential future work.

%%%%%%%%%%%%%%%%%%%%%%%%%%%%%%%%
%%%%%%%%%%%%%%%%%%%%%%%%%%%%%%%%
%%%%%%%%%%%%%%%%%%%%%%%%%%%%%%%%
%%%%%%%%%%%%%%%%%%%%%%%%%%%%%%%%

\section{Method}\label{sec:method}

We assume we have multilocus genotype data from a sample of admixed individuals arising from a number of populations.  For simplicity, we suppose that there are $N$ haploid individuals genotyped at $L$ loci, and  we denote by $X=(\xil)_{1\leq i\leq N,1\leq l\leq L}$ the observed data, where  $\xil$ is the allele of individual $i$ at locus $l$.  

\subsection{Model specification}

Let $K$ be the number of ancestral populations. We denote by $\Qi=(\qik)_{1\leq k \leq K}$ the  
vector of admixture proportions of individual $i$, where $\qik$ denotes the proportion of the genome of individual $i$ which can be traced to population $k$.  While previous works used a finite value for $K$, we will take a Bayesian nonparametric approach and let $K\rightarrow \infty$, so that there is an unbounded number of potential populations in the model. To account for dependence among loci, we use the model of \textit{linkage}  proposed by Falush et al.\ (2003)~\cite{Fal(03)}.  This  employs a hidden Markov model which splits the genome into contiguous chunks with common ancestry.  The model is parameterised by: $\dl$, the genetic distance between locus $l$ and locus $l+1$, for each $l=1,\ldots,L-1$, and $r$, the rate at which splits occur.  Let $\zil$ be a variable which denotes the population ancestry at locus $l$ of individual $i$, and  $\sil$ be a binary variable which denotes whether locus $l$ and locus $l+1$ are in the same chunk ($s_{il}=1$) or not ($s_{il}=0$).  The variables $S_{il}$ can be thought of as linkage indicators. The transition model is as follows:
\begin{align}
 \zione &\sim \Discrete(\Qi)  \\
\sin &\sim \Bernoulli(e^{-r\dl}), \quad l=1,\ldots,L-1 \nonumber \\
 \zin | \sin,\zil & \begin{cases}
= \zil & \text{if $\sin=1$,} \\
\sim \Discrete(\Qi) &\text{if $\sin=0$.}
\end{cases}               \nonumber
\label{eq:zmarkov}
\end{align}
The probability of a split between loci $l$ and $l+1$ is $1-e^{-r\dl}$, and the ancestral populations of each chromosome segment are independent and identically distributed, with probability $\qik$ for the ancestral population to be $k$ for each chunk.  The admixture model of Prichard et al.\ (2000)~\cite{Pri(00)} can be recovered as $r\rightarrow\infty$, as all loci become independent and the chunks consist of a single locus.  

The model is completed by specifying the likelihood function for the observed  alleles.  
We will assume that within each population Hardy-Weinberg equilibrium holds, and we can model the the allelic profile of the $k$th population simply by specifying the vector of allele frequencies, that is $\thetak=(\thetakla)_{1\le l\le L,1\le a\le A}$, where $\thetakla$ is the probability for allele $a$ at locus $l$ in population $k$.  That is,
\begin{align}
\xil|\zil=k &\sim \Discrete(\thetakl)
\end{align}
where $\thetakl=(\thetakla)_{1\le a\le A}$.  For example, in the case of single nucleotide polymorphism (SNP) data, $x_{il}$ are binary valued and modelled using Bernoulli distributions with means given by $\theta_{kl1}$.

\subsection{Prior specification and Bayesian nonparametrics}

We use a Dirichlet prior for the admixture proportions $\Qi$.  The typical prior in previous works \cite{Bal(95),Ran(97),Pri(00),Fal(03)} is given by a symmetric Dirichlet distribution, which assumes  that all populations have a priori equal contribution to the observed genomes.  We will use an asymmetric Dirichlet with mean $\Qz=(\qzk)_{1\le k\le K}$ instead, to capture the assumption that some populations may be more prevalent than others, so has  \emph{a priori} higher chance of contributing more genetic material to each individual (see also Anderson \cite{And(01)} and Anderson and Thompson \cite{And(02)}):
\begin{align}
\Qi | \Qz &\sim \Dirichlet(\alpha \Qz) \label{eq:Qi}
\end{align}
where $\alpha>0$ is a parameter which controls the concentration of the Dirichlet prior around its mean $\Qz$.  

The asymmetric Dirichlet also allows for a Bayesian nonparametric model, in which the number of populations $K$ is taken to be infinite, while the corresponding infinite $K$ limit does not lead to a mathematically well-defined model for the symmetric Dirichlet.  Specifically, consider a hierarchical prior on $\Qz$ expressed as the so-called stick-breaking representation \cite{Set(94)},
\begin{align}
&\text{For $j=1,2,\ldots$:} & v_{0j} &\sim \Bet(1,\alpha_0),  \label{eq:stick} \\
&& \qzj &= v_{0j} \prod_{j'=1}^{j-1} (1-v_{0j'})\nonumber
\end{align}
where $\alpha_0$ is a hyperparameter which controls the overall diversity of populations, with larger $\alpha_0$ corresponding to a larger number of populations with more uniform proportions.  The conditional distribution of $\Qi$ given $\Qz$ is still a Dirichlet as given in \eqref{eq:Qi}, though we need to extend the definition to one for infinite-dimensional vectors.  Specifically, a constructive definition of such an infinite-dimensional Dirichlet is given as follows:
\begin{align}
&\text{For $j=1,2,\ldots$:} & v_{ij} &\sim \Bet(\alpha v_{0j},\alpha(1-\sum_{j'=1}^j v_{0j'})) 
 \label{eq:qi}\\
&& \qij &= v_{ij} \prod_{j'=1}^{j-1} (1-v_{ij'}). \nonumber
\end{align}

While our model assumes theoretically the existence of an infinite number of populations, given a particular finite-sized dataset, only a finite (but random) number of populations will be used to model the data, and the posterior distribution over this number can be used to estimate the number of populations exhibited in the data; see Subsection~\ref{sec:estimating-number} for details.

The model is completed by specifying a prior on $\alpha$, $\alpha_0$, $r$ and  $\theta_{kl},k=1,\ldots,K;l=1,\ldots,L$. For each population $k$, we use independent Dirichlet for the allele frequencies at each locus. In the case of SNP data, this implies assuming independent Beta prior for each locus in each subpopulation. In our simulations and our application to the EDAR data, we take $\theta_{kl1} \sim \text{Beta} (c \mu_l, c(1-\mu_l))$, where $m_l$ denotes the prior mean for the allele frequency, assumed to be the same for all ancestral populations and $c$ is a concentration parameter. We choose independent Gamma priors for $\alpha$ and $\alpha_0$ for computational reasons. We specify a uniform prior on $\log r$, on a fairly large interval. Recall that $d_l$ denotes the genetic distance between adjacent markers. If it is measured in morgans, then $r$ can be interpreted as an estimate of $t$, the number of generations since the admixture event \cite{Fal(03)}. When the genetic distance between loci is not available, we can use as a proxy the physical distance measured in nucleotides. In this case $r$ be interpreted as an estimate of the product of $t$ and the recombination rate (expected number of crossovers per base pair per meiosis).

Another important issue which arises is the computational requirements for inference in a model with an infinite number of populations.  In this regard, a range of recent truncation and marginalisation techniques can be applied allowing for exact inference using finite computational resources \cite{Nea(00),Wal(07),Pap(08),Fav(13)}.  We propose a particular approach in Subsection~\ref{sec:inference}, after discussing in the next subsection the theoretical motivation for the hierarchical Dirichlet prior described in \eqref{eq:Qi} and \eqref{eq:stick}.

\subsection{Hierarchical Dirichlet processes}

The stick-breaking prior for the overall population prevalences \eqref{eq:stick} imposes a particular ordering on the populations, in which populations with higher index have \emph{a priori} lower prevalences.  This is undesirable from a modelling perspective as the induced ordering is artificial, while from a computational perspective it is also undesirable as it introduces a label switching problem into the inference, which can slow down convergence of inference algorithms \cite{Jas(05),Pap(08)}.  In this section we address this issue by developing a more abstract formalism for the model based on a construction of coupled random probability measures called the hierarchical Dirichlet process \cite{Teh(06)} (see also Teh and Jordan \cite{Teh(10)} for a more recent review).

Let $(\Theta,\Omega)$ be a measurable space.  The Dirichlet process $G_0\sim\DP(\alpha_0,H)$ is a random probability measure over $(\Theta,\Omega)$ with the property that for any measurable partition $(A_1,\ldots,A_L)$ of $\Theta$ the random probability vector $(G_0(A_1),\ldots,G_0(A_L))$ is distributed according to a Dirichlet distribution with parameters $(\alpha_0 H(A_1),\ldots,\alpha_0 H(A_L))$ \cite{Fer(73)}. The parameters of the process consist of a positive concentration parameter $\alpha_0$ and a base probability measure $H$ over $(\Theta,\Omega)$.  A variety of more constructive representations exist for the Dirichlet process, and the reader is referred to Ghoshal \cite{Ghos(10)} for a review on the DP and to  Lijoi and  Pr\"unster \cite{lp(10)} for a review of nonparametric prior distributions generalizing the DP.  

One of the noteworthy properties of the Dirichlet process is that the random probability measure $G_0$ is discrete almost surely, and can be written in the form
\begin{align}
G_0 = \sum_{k=1}^\infty \qzk \delta_{\theta_k}.
\end{align}
The atoms $(\theta_k)_{k\ge 1}$ are independent and identically distributed according to the base probability measure $H$, while the atom masses are independent of the atoms, and have distribution given by the stick-breaking representation \eqref{eq:stick} \cite{Set(94)}.  

In the context of admixture modelling, we will suppose that each atom in $G_0$ corresponds to a population with allelic frequencies parameterised by the atom, while the masses correspond to the population proportions or prevalences.  In other words, $\theta_k$ denotes the vector of the population specific allele frequencies for the $L$ loci under investigation. As each individual has its own population proportions while the collection of populations are shared across individuals, we can model this using the hierarchical Dirichlet process (HDP). For each individual $i$, let $G_i$ be an individual-specific atomic random probability measure. These measures are conditionally independent and identically distributed given a common base probability measure $G_0$:
\begin{align}
G_i |G_0 &\sim \DP(\alpha, G_0)
\end{align}
Since each atom in $G_i$ is drawn from $G_0$, the collection of atoms in $G_i$ is precisely those in $G_0$, while each $G_i$ has its own specific atom masses:
\begin{align}
G_i &= \sum_{k=1}^\infty \qik \delta_{\theta_k}
\end{align}
where the masses $(\qik)_{k\ge 1}$ have distribution as given in \eqref{eq:qi}. The HDP allows sharing of the ancestral among the indivual distributions as the $G_i$
place atoms at the same discrete locations determined by $G_0$ (see Teh et al.\ (2006)~\cite{Teh(06)} for details).  

We refer to the proposed model as HDPStructure.
In summary, 
$G_i$ describes the proportion of the alleles on $x_i=(x_{i1},\ldots,x_{iL})$ coming from each of the populations, as well as the
parameters of the populations. We model the sequence $x_i$ given $G_i$ as follows: (i) first we place segment boundaries according to an
nonhomogeneous Poisson process with rate $r d_l$, (ii) then the alleles on each segment are generated by
picking a population of origin according to $G_i$, then sampling
the alleles according to the population distribution.
We have expressed the hierarchical prior over the population proportions \eqref{eq:stick}, \eqref{eq:qi} as the joint distribution of atom masses in a HDP, while the atoms correspond to the population parameters.  Further, while the stick-breaking representation imposes a particular ordering among the atoms, there is no ordering of atoms in the representation as random probability measures themselves.  As we will see next, this allows for an efficient Markov chain Monte Carlo algorithm for posterior simulation.

\subsection{Markov Chain Monte Carlo}\label{sec:inference}

In this section we describe a Markov chain Monte Carlo (MCMC) algorithm for posterior simulation in the HDPStructure model.  The MCMC sampling algorithm iterates between updates to  the random probability measures $(G_i)_{0\le i\le N}$,  the latent state sequences $(\sil,\zil)_{1\le i\le N,1\le l < L}$, and the model parameters in turn, each update conditional upon all the other variables in the model.   Updates to the random probability measures make use of another representation of the HDP called the Chinese restaurant franchise \cite{Teh(06)}, as well as a retrospective slice sampling technique which allows for a finite truncation to the random probability measures while retaining exactness of the procedure.  Updates to the latent state sequences make use of an efficient forward filtering-backward sampling procedure as a Metropolis-Hastings proposal distribution.  Finally, updates to model parameters are straightforward one-dimensional Metropolis-Hastings updates. Detailed descriptions of these updates are included in the Appendix. MATLAB software implementing this MCMC scheme is freely available at \url{http://BigBayes.github.io/HDPStructure}.

\subsubsection{Updates to random probability measures}\label{sec:estimating-number}

Conditioned on the model parameters and latent state sequences, the update to the random probability measures $(G_i)_{0\le i\le N}$ follow standard results for the hierarchical Dirichlet process \cite{Teh(06)}.  As noted previously, since the data is finite, the number of populations used to model the data is finite as well.  Conditioned on the latent state sequences $\{ z_{il}\}$, suppose the number of such populations (as a function of the latent state sequences) is $K^*$.  For simplicity, we may index these populations as $1,\ldots, K^*$.  The random probability measures can be expressed as:
\begin{align}
G_0 &= \sum_{k=1}^{K^*} q_{0k} \delta_{\theta_k} + w_0 G_0' &
G_i &= \sum_{k=1}^{K^*} q_{ik} \delta_{\theta_k} + w_i G_i'
\end{align}
for each $i=1,\ldots,N$, where $w_i$ is the total mass of all other atoms in $G_i$, which are collected, after normalising by $w_i$, in a random probability measure $G_i'$.  

For each $i=1,\ldots,N$ and $k=1,\ldots,K^*$, let $n_{ik}$ be the number of DNA segments in sequence $i$ assigned to population $k$.  In the Chinese restaurant franchise representation of the HDP, we introduce a set of discrete auxiliary variables $m_{ik}$, taking value $0$ if $n_{ik}=0$, and values in the range $\{1,\ldots,n_{ik}\}$ when $n_{ik}>0$.  Define $n_{0k}=\sum_{i=1}^N m_{ik}$.  Then the conditional distributions of the random probability measures given $(n_{ik},m_{ik})_{0\le i\le N,1\le k\le K^*}$ is described by the following \cite{Teh(06)},
\begin{align}
(q_{01},\ldots,q_{0K^*},w_0) | (n_{ik},m_{ik})
&\sim \Dirichlet(n_{01},\ldots,n_{0K^*},\alpha_0) \\
(q_{i1},\ldots,q_{iK^*},w_i) | (n_{ik},m_{ik}), (q_{01},\ldots,q_{0K^*}),w_0
&\sim \Dirichlet(\alpha q_{01} + n_{i1},\ldots,\alpha q_{0K^*}+n_{0K^*},\alpha w_0) \nonumber \\
G_0' | (n_{ik},m_{ik})&\sim \DP(\alpha_0,H) \\
G_i' | (n_{ik},m_{ik}), G_0' &\sim \DP(\alpha, G_0') \nonumber
\end{align}
where the masses form a hierarchy of finite-dimensional Dirichlet distributions while the random probability measures are independent of the masses and form a hierarchy of DPs as in the prior.

A final point of consideration relates to the fact that the random probability measures $G_0', (G_i')$ have infinitely many atoms, so not all can be simulated explicitly with finite computational resources.
We address this using a retrospective slice sampling technique to truncate the random probability measures while retaining exactness \cite{Wal(07),Pap(08),Gri(13)}.  For each individual $i$, we introduce an auxiliary slice variable $C_i$, with conditional distribution given the other variables in the model:
\begin{align}
q^\text{min}_i  & = \min_{l=1,\ldots,L} q_{iz_{il}} \label{eq:qmin}\\
C_i | (n_{ik},m_{ik}), G_0, (G_i) &\sim \Uniform[0,q^\text{min}_i].
\end{align}
The slice variables are sampled just before the latent state sequences (whose updates are described in the next subsection).  Further, conditioned on the slice variables, populations whose mass fall below $C_i$ will have zero probability to be selected when the latent state sequence for individual $i$ is updated.  As a consequence, only the (finitely many) atoms with mass above the minimum threshold $\min_i C_i$ need be simulated.  This can be achieved by simulating $G_0'$ and $(G_i')$ using the hierarchical stick-breaking representation \eqref{eq:stick}, \eqref{eq:qi} until the left-over mass falls below the threshold.  

\subsubsection{Updates to latent state sequences}

We will use a forward-filtering backward-sampling algorithm to resample the latent state sequences one at a time.  Conditioned on the slice variable $C_i$, only populations with $q_{ik}>C_i$ will have positive probability of being selected, and so the forward-backward algorithm can be computationally tractable.  However, as the slice variable depends on all latent state variables, conditioning on the slice variable introduces complex dependencies among the latent state variables which precludes an exact and efficient forward filtering algorithm.  We propose instead to ignore the dependencies caused by the slice variable, and use the resulting efficient forward-backward algorithm as a Metropolis-Hastings proposal.

Suppose that there are $K_i$ populations with proportions above the slice threshold $C_i$. For simplicity of exposition, we will reindex the populations such that their indices are simply $\{1,\ldots,K_i\}$.  The forward-backward algorithm efficiently samples from the following proposal distribution which ignores the slice threshold $C_i$:
\begin{align}
Q((z_{il},s_{il})_{l=1}^L) \propto \Prob((z_{il},s_{il})_{l=1}^L,G_i)\Prob((x_{il})_{l=1}^L|(z_{il},s_{il})_{l=1}^L|G_i)
\end{align}
where the population indicators range only over $1,\ldots,K_i$.

The forward filtering phase first computes the following probabilities using dynamic programming:
\begin{align}
M^{il}_{bk} &= \Prob(x_{i1},\ldots,x_{il},s_{il}=b,z_{il}=k|(q_{ik},\theta_k)_{k=1}^{K_i})\\
M^{il}_{\cdot k} &= \Prob(x_{i1},\ldots,x_{il},z_{il}=k|(q_{ik},\theta_k)_{k=1}^{K_i})\nonumber \\
M^{il}_{\cdot \cdot} &= \Prob(x_{i1},\ldots,x_{il}|(q_{ik},\theta_k)_{k=1}^{K_i}) \nonumber
\end{align}
with $b \in \{0,1\}$ and $k\in \{1,\ldots, K_i\}$.
The dynamic programme starts at $l=1$:
\begin{align*}
M^{i1}_{\cdot k} &= \theta_{k1x_{i1}} q_{ik}\\
\intertext{and proceeds with $l=2,\ldots,L$:}
M^{il}_{1k} &= \theta_{klx_{il}} e^{-rd_{l-1}} M^{il-1}_{\cdot k} \nonumber &
M^{il}_{\cdot k} &= M^{il}_{0k} + M^{il}_{1k} \nonumber\\
M^{il}_{0k} &= \theta_{klx_{il}} (1-e^{-rd_{l-1}})  q_{ik} M^{il-1}_{\cdot \cdot} \nonumber&
M^{il}_{\cdot \cdot} &= \sum_{k=1}^{K_i} M^{il}_{\cdot k}. \nonumber
\end{align*}
Recall that $\theta_{klx_{il} }$ is the probability that locus $l$ in population $k$ assume the observed value $x_{il}$. 
The backward phase samples from the proposal distribution, starting at  $l=L$:
\begin{align*}
Q(z_{iL}=k) &\propto M^{iL}_{\cdot k} \\
Q(s_{iL}=b|z_{iL}=k) &\propto  M^{iL}_{bk} \\
\intertext{and iterates backwards, for $l=L-1,\ldots,1$:}
Q(z_{il}=k|s_{il+1}=1,z_{il+1}=k') &\propto \boldsymbol{1}(k=k') \\
Q(z_{il}=k|s_{il+1}=0,z_{il+1}=k') &\propto M^{il}_{\cdot k} \\
Q(s_{il}=b|z_{il}=k) &\propto M^{il}_{bk}.
\end{align*}
where $\boldsymbol{1}(\cdot)$ denotes the indicator function and $s_{i1}=0$ by construction. In this way we obtain a new sample for the collection of $s_{il}$ and $z_{il}$.
Finally, the Metropolis-Hastings acceptance probability is a simple expression which accounts for the effect of conditioning on $C_i$:
\begin{align}
\min\left(1,\frac{q_i^\text{min-cur}}{q_i^{\text{min-prop}}}\right)
\end{align}
where $q_i^\text{min-cur}$ and $q_i^{\text{min-prop}}$ indicate the minimum population proportions \eqref{eq:qmin} under the current  and proposed states respectively.

The forward-backward algorithm has a computational scaling of $\mathcal{O}(LK_i)$, linear in both the length of the sequence and the number of potential populations, and is the most computationally expensive part of the MCMC algorithm. It must be noted that, since the $(G_i)$ are conditionally independent given $G_0$, the algorithm  can be easily parallelised  so as to exploit modern parallel computation technology.

\subsection{Extensions}
The model can be straightforwardly extended to diploid or polyploid data, by assuming that, for each individual $i$, the $z_i$ along 
each of individual $i$'s chromosomes form independent 
Markov chains satisfying \eqref{eq:zmarkov}.
Other extensions of the Bayesian nonparametric admixture model can be introduced to allow correlated allele frequencies. For instance, following the approaches of Pritchard et al.\ (2000)~\cite{Pri(00)} and Falush et al.\ (2003)~\cite{Fal(03)}, it is straightforward to introduce a Bayesian nonparametric admixture model with correlated allele frequencies. Specifically, we can assume that  allele frequencies in one population provide information about the allele frequencies in another population, i.e.\ frequencies in the different populations are likely to be similar (probably due to migration or shared
ancestry). This can be achieved by specifying a more complex prior structure   on $\theta_{kl.}$, for example employing the correlated allele frequencies model of Falush et al.\ (2003)~\cite{Fal(03)}, which assumes that allele frequencies at locus $l$ in different populations are deviations from allele frequencies in a hypothetical ancestral population.   

At the moment, we use only genetic data to infer admixture parameters. Often it can be useful to include in the model extra information such as physical characteristics (e.g.\ ethnicity) of sampled individuals or
geographic sampling locations \cite{Hub(09)}. These new sources of information would modify the clustering structure and would allow the proportion
of individuals assigned to a particular cluster to depend on the new information. This would require a specification of a spatiality dependent model on the weights of the random measures in the HDP. 

From a Bayesian parametric perspective, we could also employ  other priors such as the Pitman-Yor process \cite{Pit(97)} and the hierarchical Pitman-Yor process \cite{Teh(10)}. The Pitman-Yor process is a two-parameter generalization of the DP, for which  a stick-breaking construction and a Chinese restaurant representation still hold.  Under certain assumptions, it can be shown that in the Pitman-Yor process the number of clusters grows much faster than for a standard DP and that  the cluster sizes decay according to a power law. This property makes the   Pitman-Yor process a more suitable choice in many applications. The implementation of this more flexible prior would require more expensive computations due to the larger number of extant populations possible.

%%%%%%%%%%%%%%%%%%%%%%%%%%%%%%%%
%%%%%%%%%%%%%%%%%%%%%%%%%%%%%%%%
%%%%%%%%%%%%%%%%%%%%%%%%%%%%%%%%
%%%%%%%%%%%%%%%%%%%%%%%%%%%%%%%%

\section{Simulation studies} \label{sec:simulation}

\begin{center}
\begin{figure}
\label{fig:postK}
\includegraphics[width=\textwidth]{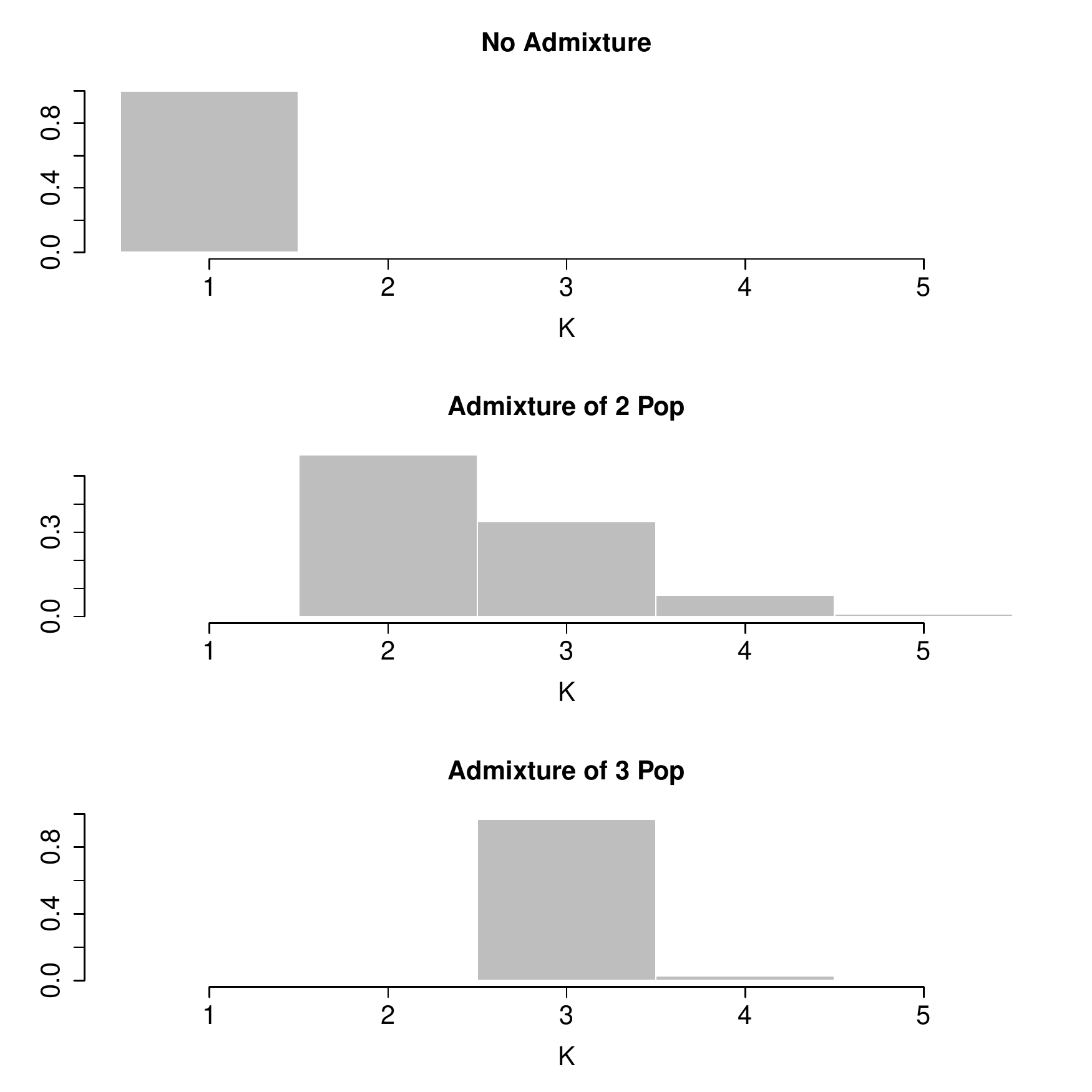}
\caption{$\Pr(K\mid \text{Data})$ under the three simulated scenarios.}
\end{figure}
\end{center}

In order to assess the performance of the model in recovering the 
number of ancestral populations, we perform simulations based on three demographic scenarios. Each scenario consists of 200 haploid sequences. We consider 60 bi-allelic genetic markers on a 100Kbp segment. Coalescent simulations were performed  using the software {\tt ms} \cite{Hud(09)}. Mutation and recombination rates were set to $2\times 10^{-8}  $ and $10^{-8}$ per base pair per generation respectively. The focus is to identify the populations of origin and to infer demographic history. In particular, the main goal is inference on $K$. The three simulated scenarios we consider are: (i) sample from a single random mating population; (ii) admixture model with two parental populations and admixture proportion of 0.6 and 0.4; (iii) admixture model with three parental populations and admixture proportion 0.5, 0.4 and 0.1. Figure \ref{fig:postK} shows the posterior distribution of $K$ (after burn-in) under the three scenarios. HDPStructure successfully recovers the number of ancestral populations.   

In the simulations we have set the parameters for the Gamma priors on $\alpha$ and $\alpha_0$ as follows:  $\alpha\sim \text{Gamma}(1,1)$ and  $\alpha_0\sim \text{Gamma}(5,1)$. Although inference can be sensitive to the choice of the prior on the number of populations, i.e. to the prior specification on $\alpha$ and $\alpha_0$, we note that as the number of sequences and/or markers increases the model tends to generate spurious clusters, i.e.\ clusters with very few individuals in them. This is in line with recent results on the clustering properties of the Dirichlet Process \citep{Mill(14)}. Nevertheless, the number of clusters that covers the majority of the data, i.e. 95-99\%, is quite robust to prior specifications (sensitivity analysis results not shown).  In general, the biological interpretation of $K$ is difficult. This is in agreement with the suggestion of Pritchard et al.\ (2010)\cite{Pri(10)}:

\begin{quotation}
We may not be able to know the \textbf{TRUE} value of $K$, but we should aim for the smallest value of $K$ that captures the major structure in the data.
\end{quotation}       

We compare our results with the software STRUCTURE \citep{Pri(00),Fal(03),Hub(09)}, which is arguably the most widely used software in applications to infer population structure. This software implements the parametric version of our model, with fixed $K$ 
(\url{http://pritchardlab.stanford.edu/structure.html}). We have run the  Linkage model described in Falush et al.~(2003) \cite{Fal(03)}, using default settings and assuming independent allele frequencies between markers. To estimate $K$, Pritchard et al.\ (2000)~\cite{Pri(00)} suggested using the value of $K$ which maximises the estimated model log-likelihood, $ \log \Pr(\text{Data}\mid K)$.  This latter quantity is estimated by the MCMC run using an approximation based on the harmonic mean estimator of the Bayesian deviance. For each scenario, we have run STRUCTURE for each value of $K$, $K=1,\ldots,8$. Figure \label{fig:structres} shows the estimated  $ \log \Pr(\text{Data}\mid K)$ for the three simulated examples. The  value $K=1$ seems to maximises the model log-likelihood under all scenarios.  In general, the authors of the software warn against possible drawback of using this criterion and to interpret the results with caution and give suggestions for improvement.

%%%%%%%%%%%%%%%%%%%%%%%%%%%%%%%%
%%%%%%%%%%%%%%%%%%%%%%%%%%%%%%%%
%%%%%%%%%%%%%%%%%%%%%%%%%%%%%%%%
%%%%%%%%%%%%%%%%%%%%%%%%%%%%%%%%

\section{Experiments}

\label{sec:experiments}
We demonstrate our model on a dataset of 372 Colombians recently genotyped on the Illumina Human610-Quadv1\_B SNP array as part of a genome-wide association study \cite{Sch(12)}. Latin American samples are uniquely advantageous for this purpose \cite{Ruiz(14)} because of their well-documented history of extensive mixing between Native Americans and people arriving from Europe and Africa. This continental admixture, which has occurred for the past 500 years (or about 20-25 generations), gives rise to haplotype blocks which are about the right length for such analysis. Ancient admixture produces very short haplotype fragments which are hard to assign ancestry with certainty, while very recent admixture allows only large haplotype blocks and there is not sufficient variation in ancestry for individuals.

\begin{center}
\begin{figure}
\label{fig:LD}
\includegraphics[width=\textwidth]{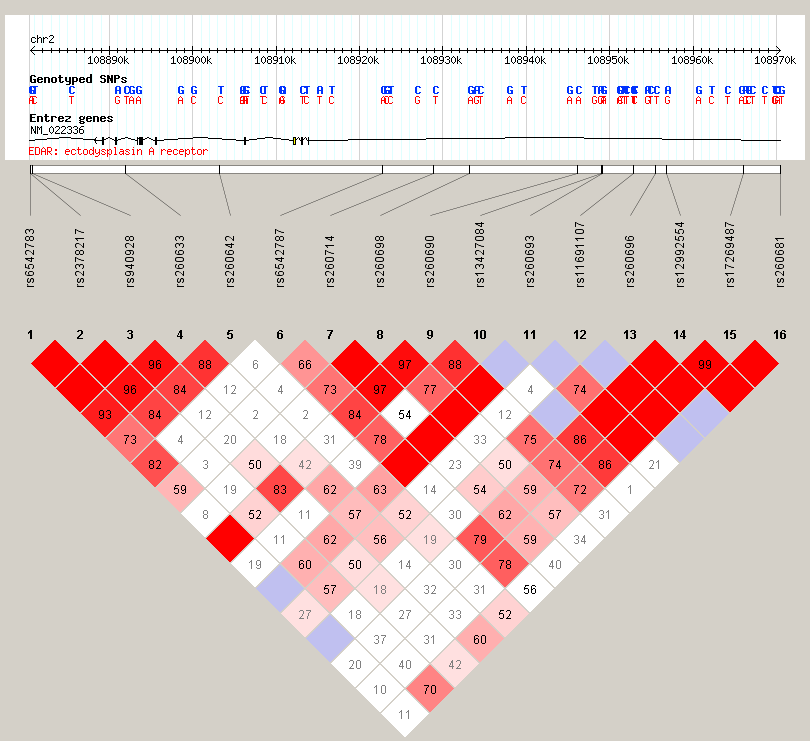}
\caption{LD plot for the EDAR region in the Colombian samples.}
\end{figure}
\end{center}

The Native American population arose as a branch of the East Asian populations who were separated over 15,000 years ago and consequently isolation and genetic drift shaped their genetic landscape. This caused many SNPs to drift even more than their East Asian counterparts, eventually becoming fixed at the alternative allele. The Ectodysplasin-A receptor (EDAR) gene, located on chromosome 2, is a common example, in particular SNP rs3827760 \cite{Mik(09)}, whose ancestral A allele is 100\% prevalent in European and African populations, but the alternative G allele is seen at 94\% frequency in Han Chinese and 100\% in Native Americans. The SNP, a missense mutation, has been observed to have a range of functional effects in humans and replicated in other mammals such as mice, including the characteristic straight hair shape in East Asians \cite{Fuj(08),Tan(13)} and dental morphology \cite{Kim(09),Par(12)}. Our dataset 
does not contain rs3827760, but  neighbouring SNPs in LD to rs3827760 are included in the chip panel.
This shows a strength of  our model does as we manage  to capture the ancestries even in absence of SNP rs3827760, the well-known causal and ancestry-informative SNP, by making good use of LD information. 
Figure \ref{fig:LD} shows the LD plot for the EDAR region in the Colombian samples.
Overall, EDAR signalling acts during prenatal development to specify the location, size and shape of ectodermal appendages, such as hair follicles, teeth and glands \cite{Mik(09)}. Therefore, we considered EDAR to be an interesting candidate for admixture analysis as it carries information regarding ancestry due to its variation across ancestry as well as its range of functional effects, which means it may be showing some signal of selection.

Genotype information on 372 individuals for 16 SNPs in the EDAR region was available from our Illumina chip data. Genotypes were phased for conversion to haplotype format by ShapeIt2~\cite{Del(13)}. Data from a total of 828 individuals sampled in putative parental populations were used as reference ancestral groups. These were selected from HAPMAP, the CEPH-HDGP cell panel \cite{Li(08)} and from published Native American data \cite{Rei(12)} as follows: 169 Africans (from 5 populations from Sub-Saharan West Africa), 299 Europeans (from 7 West and South European populations) and 360 Native Americans (from 47 populations from México Southwards).

We ran the MCMC sampler for 50,000 iterations. We collected samples after a burn-in of 20,0000 iterations and thinned every 30 iterations. We specify the following prior distributions for the precision parameters in the HDP:  $a_0\sim \text{Gamma}(1,1),\alpha\sim \text{Gamma}(10,20)$. We centre the prior for the mean parameters of the Beta base measure of the HDP around the overall observed allele frequencies, with $c=0.01$.  The prior for $\log r$ is a Uniform on the interval $[-500,5]$.

\begin{center}
\begin{figure}
\label{fig:K}
%\begin{tabular}{cc}
%\includegraphics[width=0.5\textwidth]{rawdata}&
%\includegraphics[width=0.5\textwidth]{modeclus}
%\end{tabular}
\includegraphics[width=\textwidth]{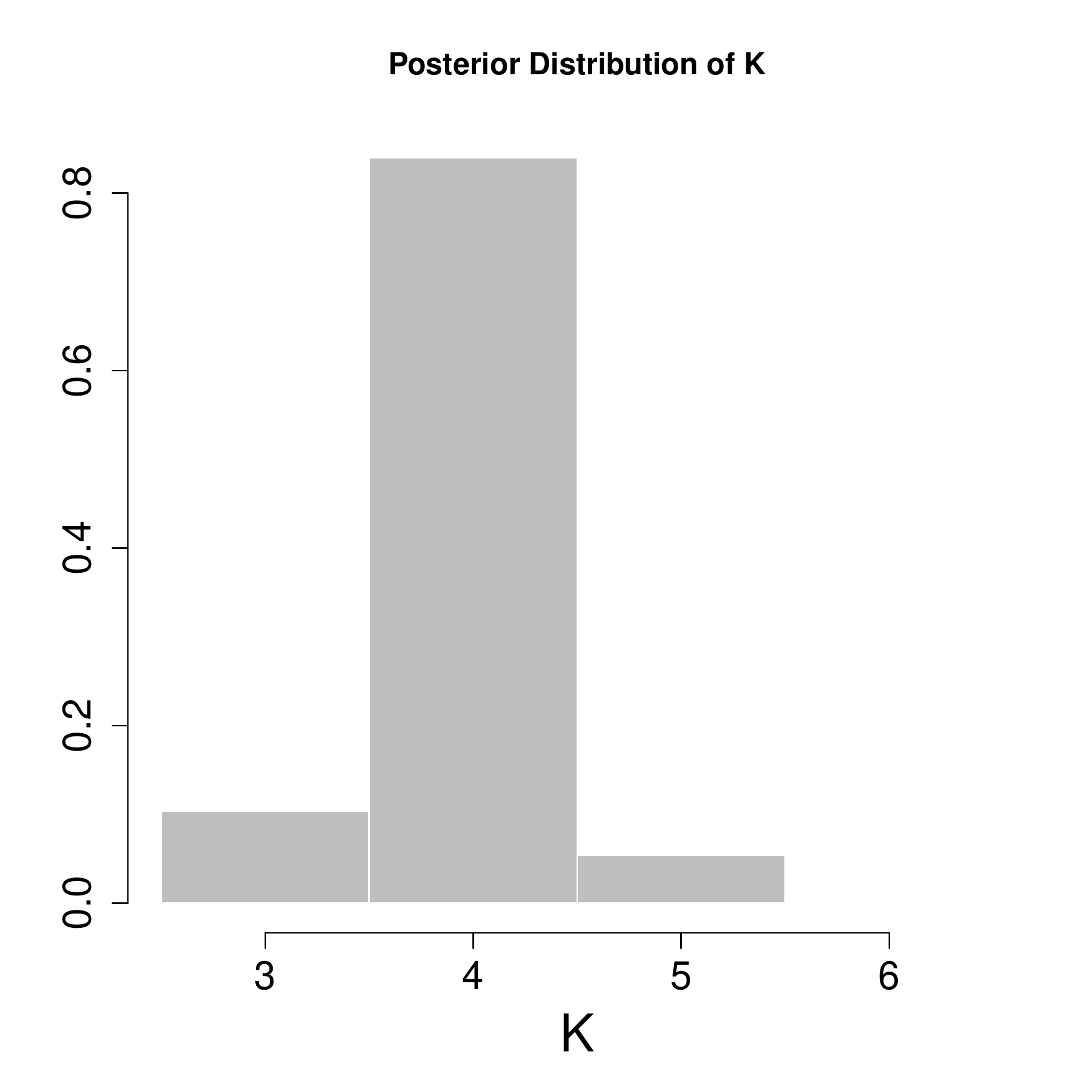}
\caption{EDAR data: posterior distribution of the number of clusters $K$.}
\end{figure}
\end{center}

\begin{center}
\begin{figure}
\label{fig:clus}
\begin{tabular}{c}
\includegraphics[width=\textwidth]{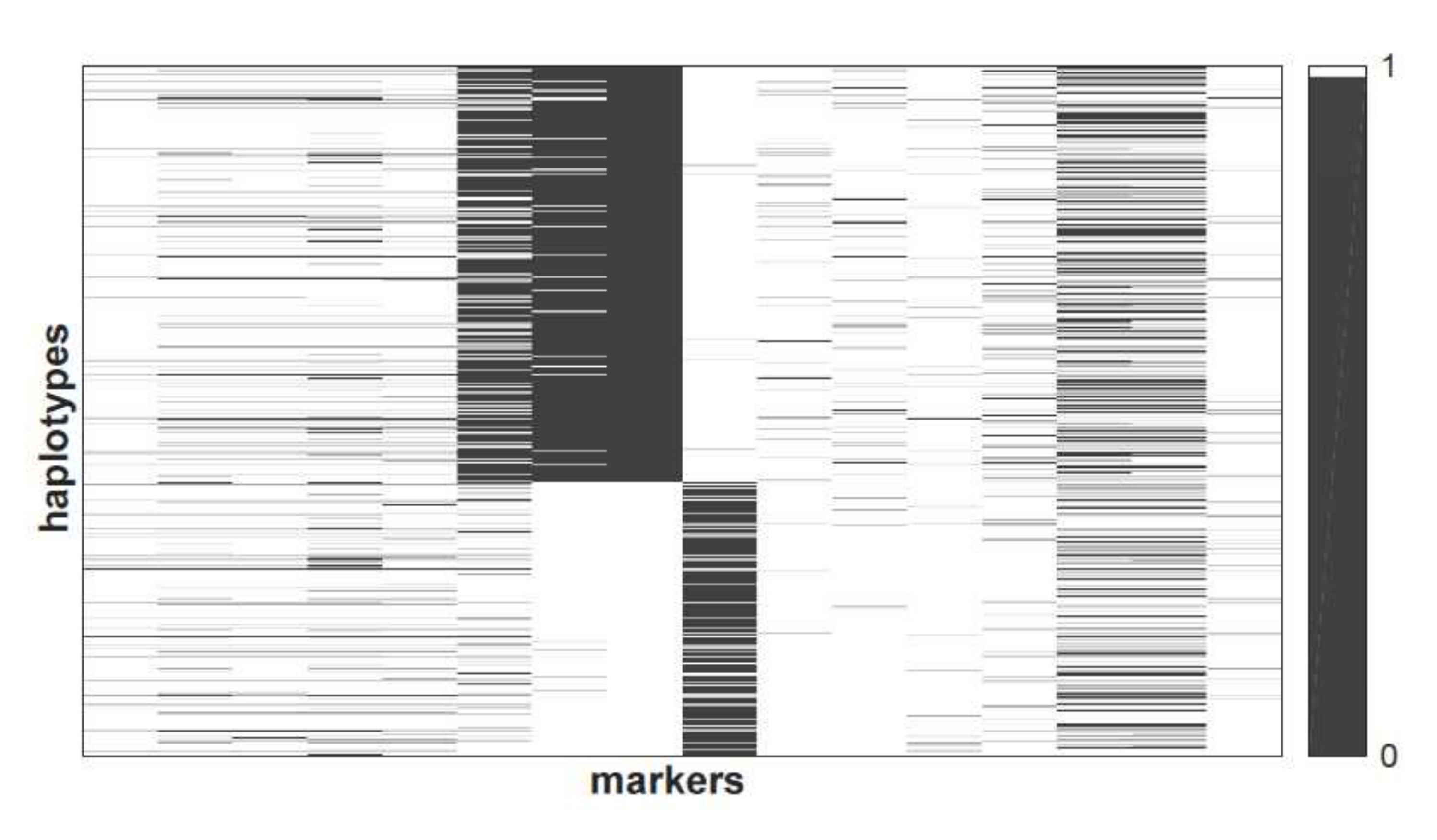}\\
\includegraphics[width=\textwidth]{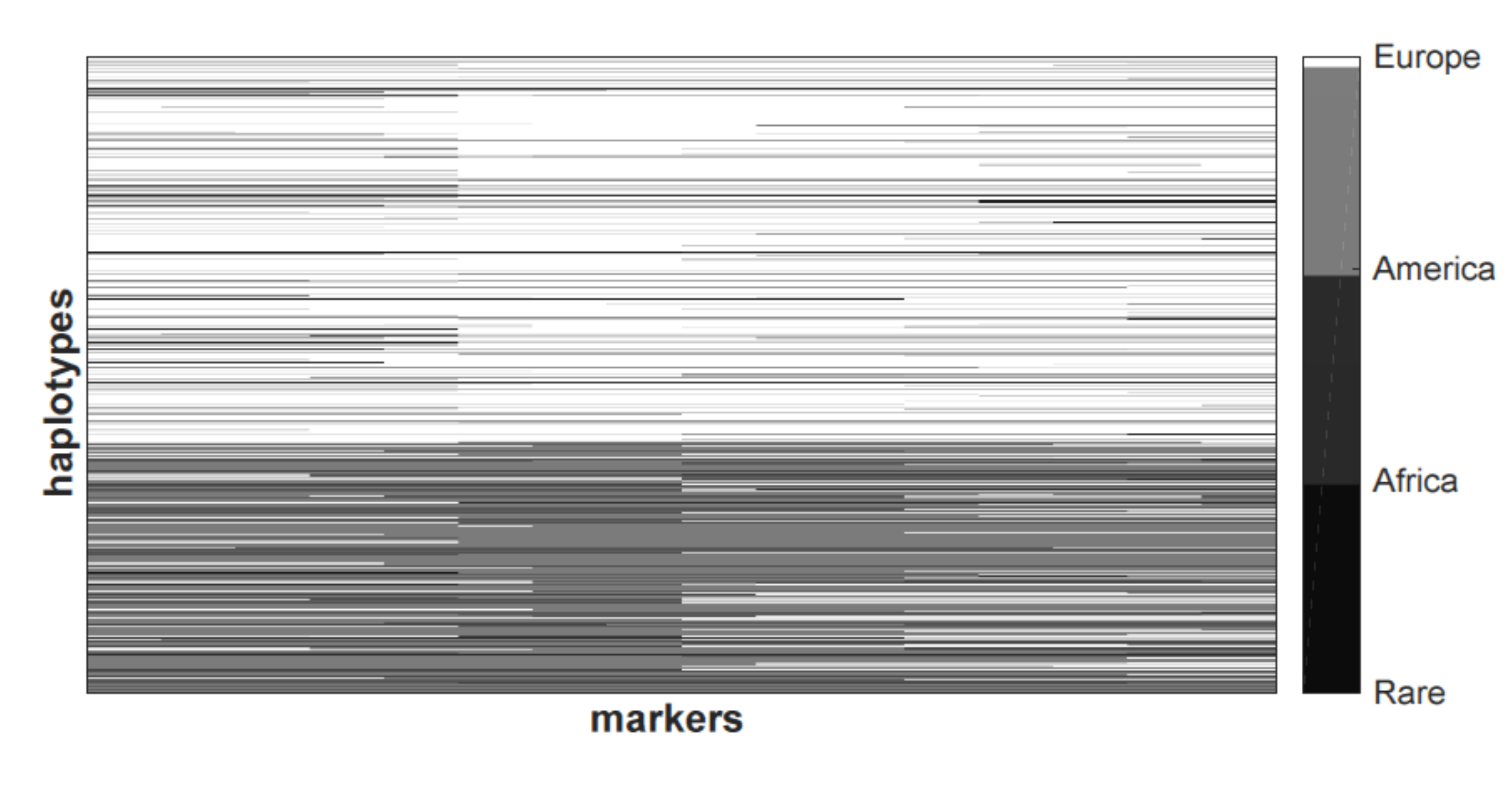}
\end{tabular}
\caption{The top panel shows the genetic data, with black representing the 0 allele for each SNP. The bottom panel presents summary of the posterior population assignment obtained by minimising the Binder loss function.}
\end{figure}
\end{center}

\begin{center}
\begin{figure}
\label{fig:dataclus}
\begin{tabular}{cc}
\includegraphics[width=0.5\textwidth]{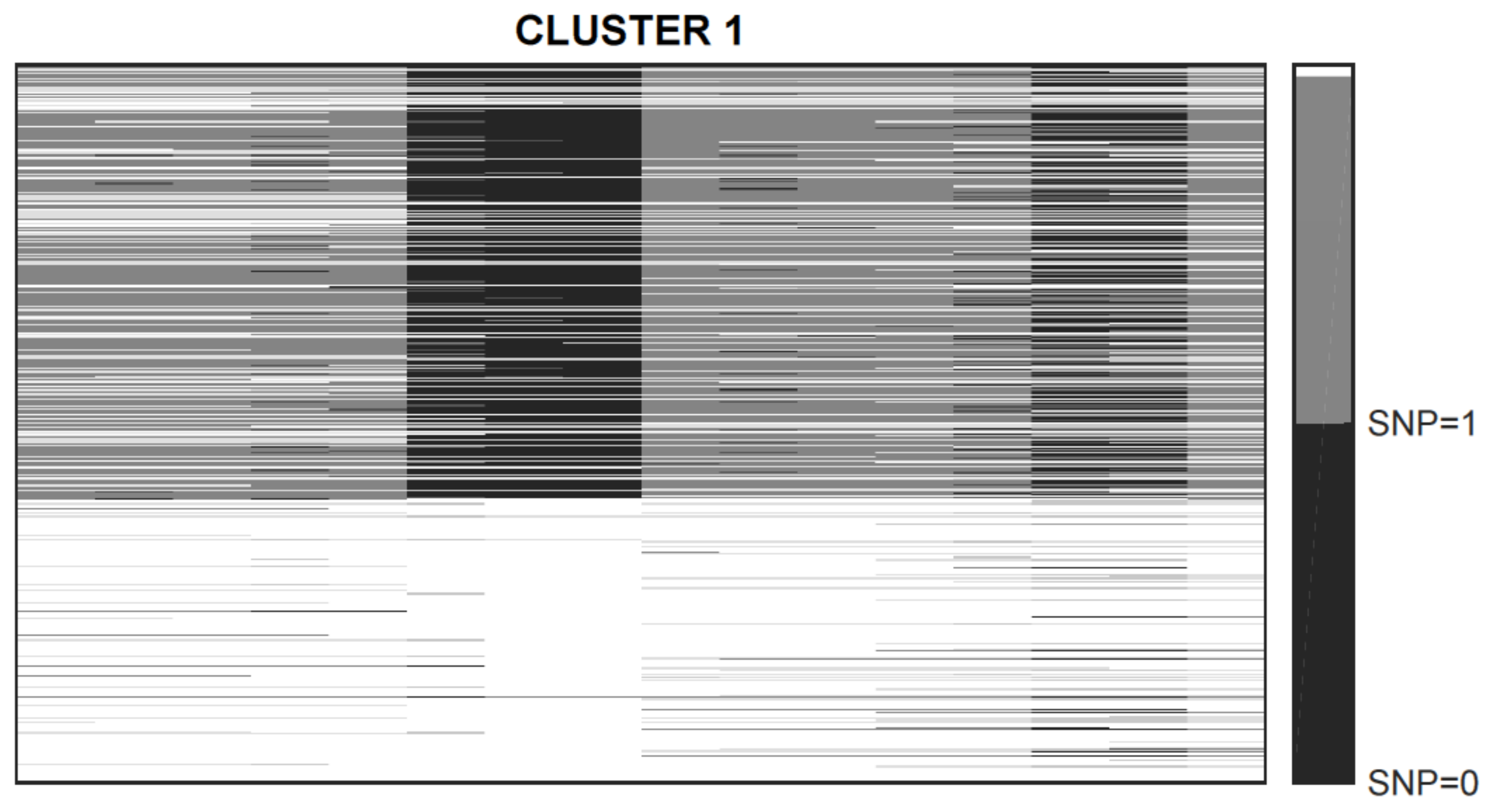} &
\includegraphics[width=0.5\textwidth]{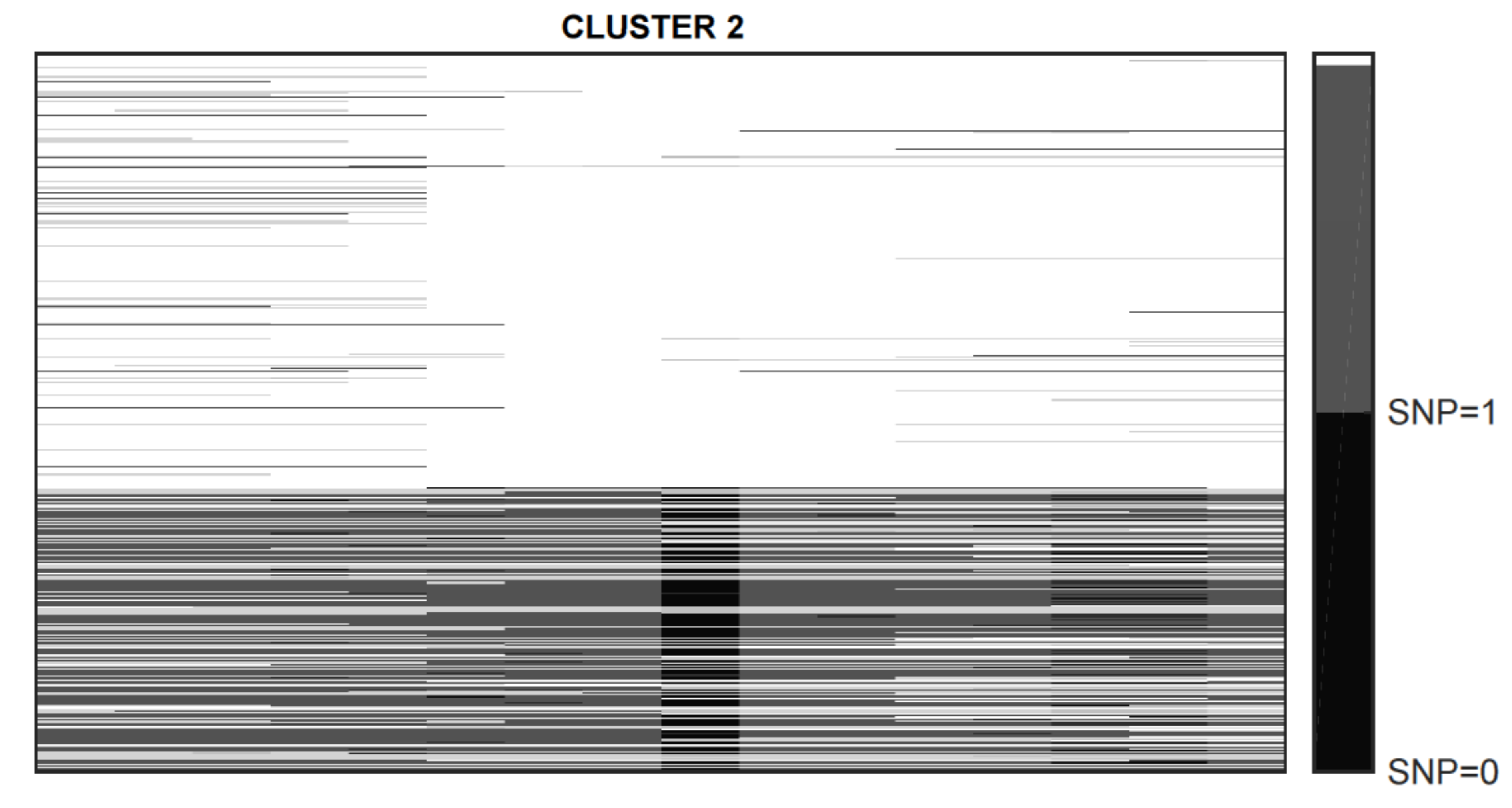} \\
\includegraphics[width=0.5\textwidth]{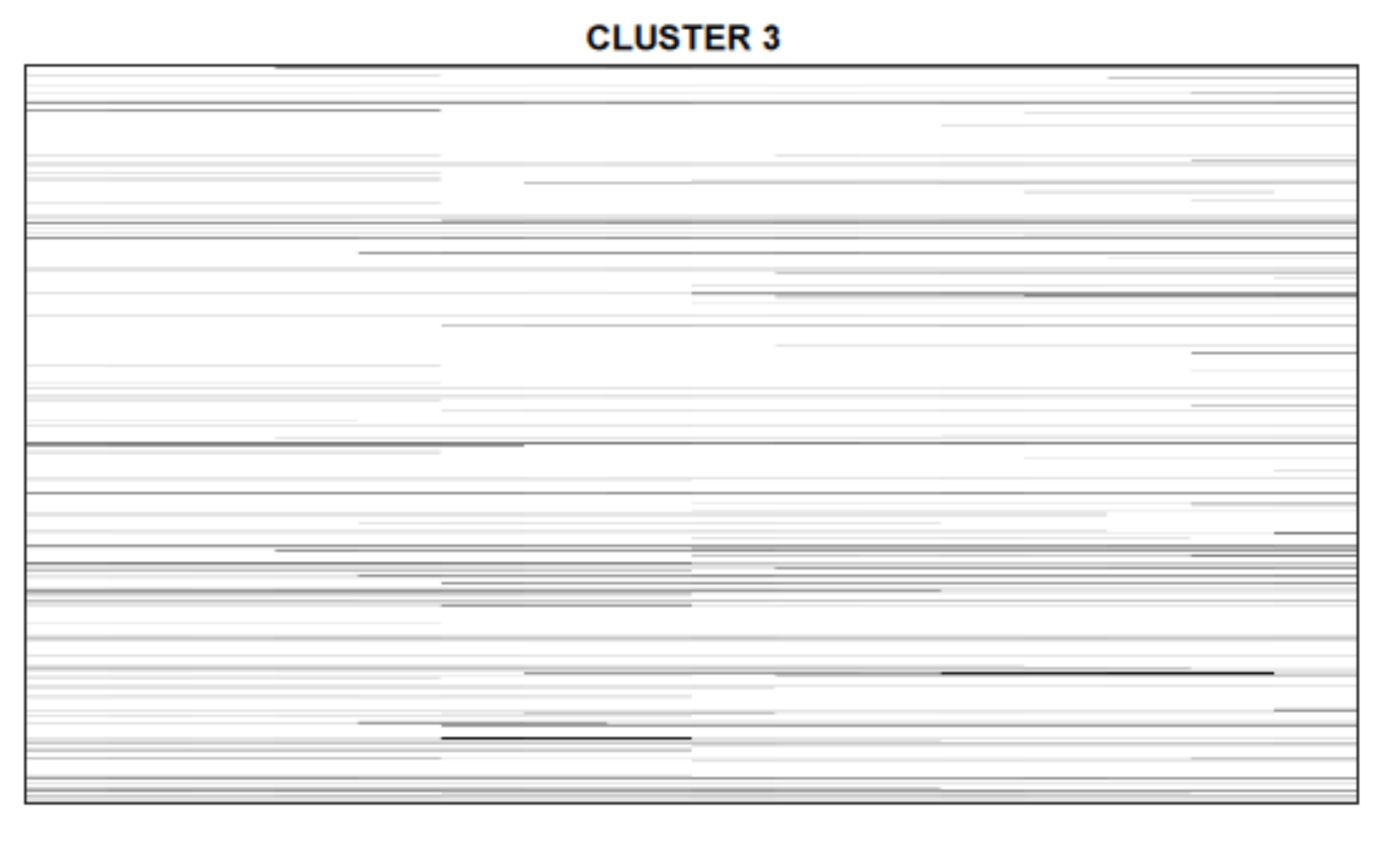} &
\includegraphics[width=0.5\textwidth]{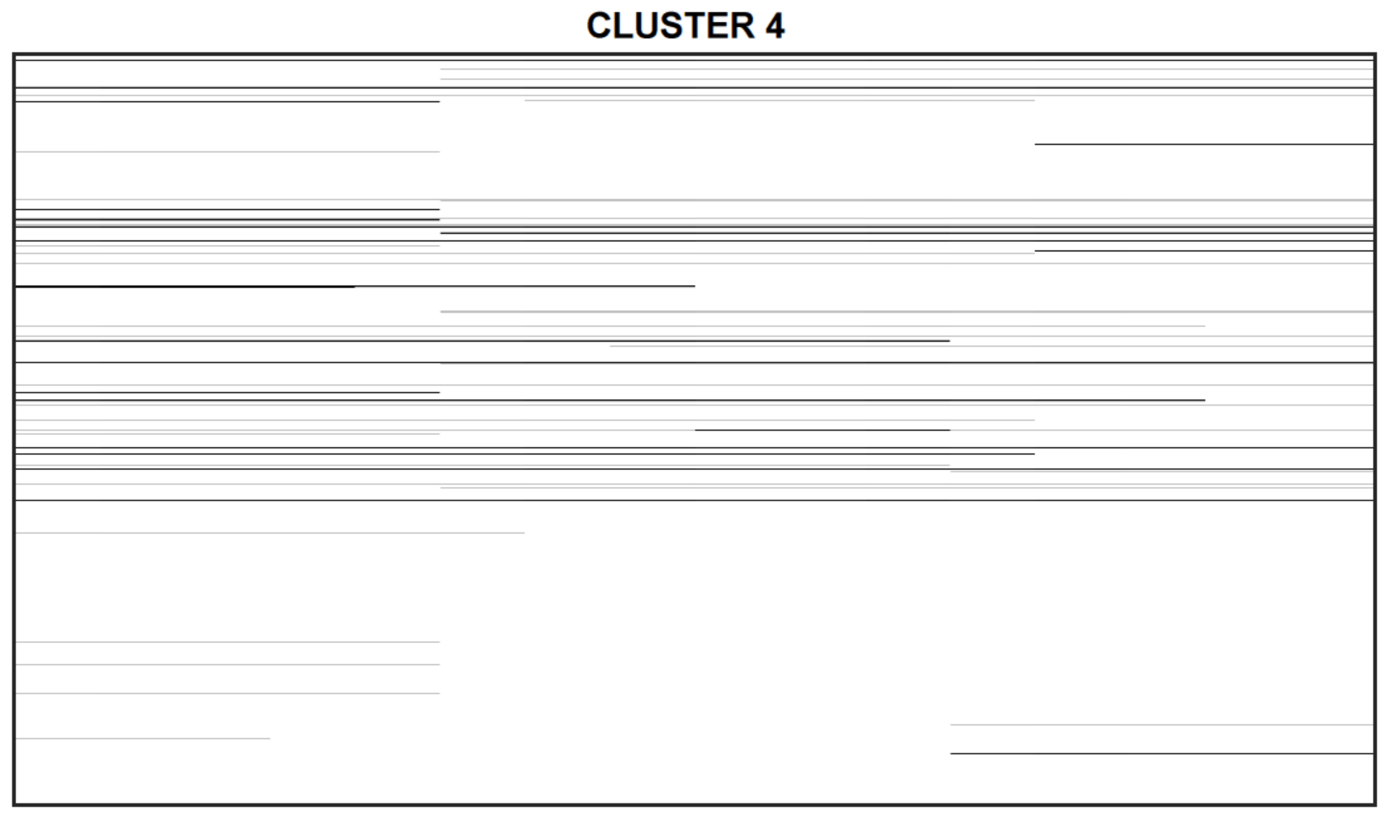}
\end{tabular}
\caption{Each panel shows the data distribution for each of the four major populations: white indicates markers on each sequences not assigned to the specific population, grey denotes the 1 allele and black denotes the 0 allele.}
\end{figure}
\end{center}

\begin{center}
\begin{figure}
\label{fig:rareclus}
\includegraphics[width=\textwidth]{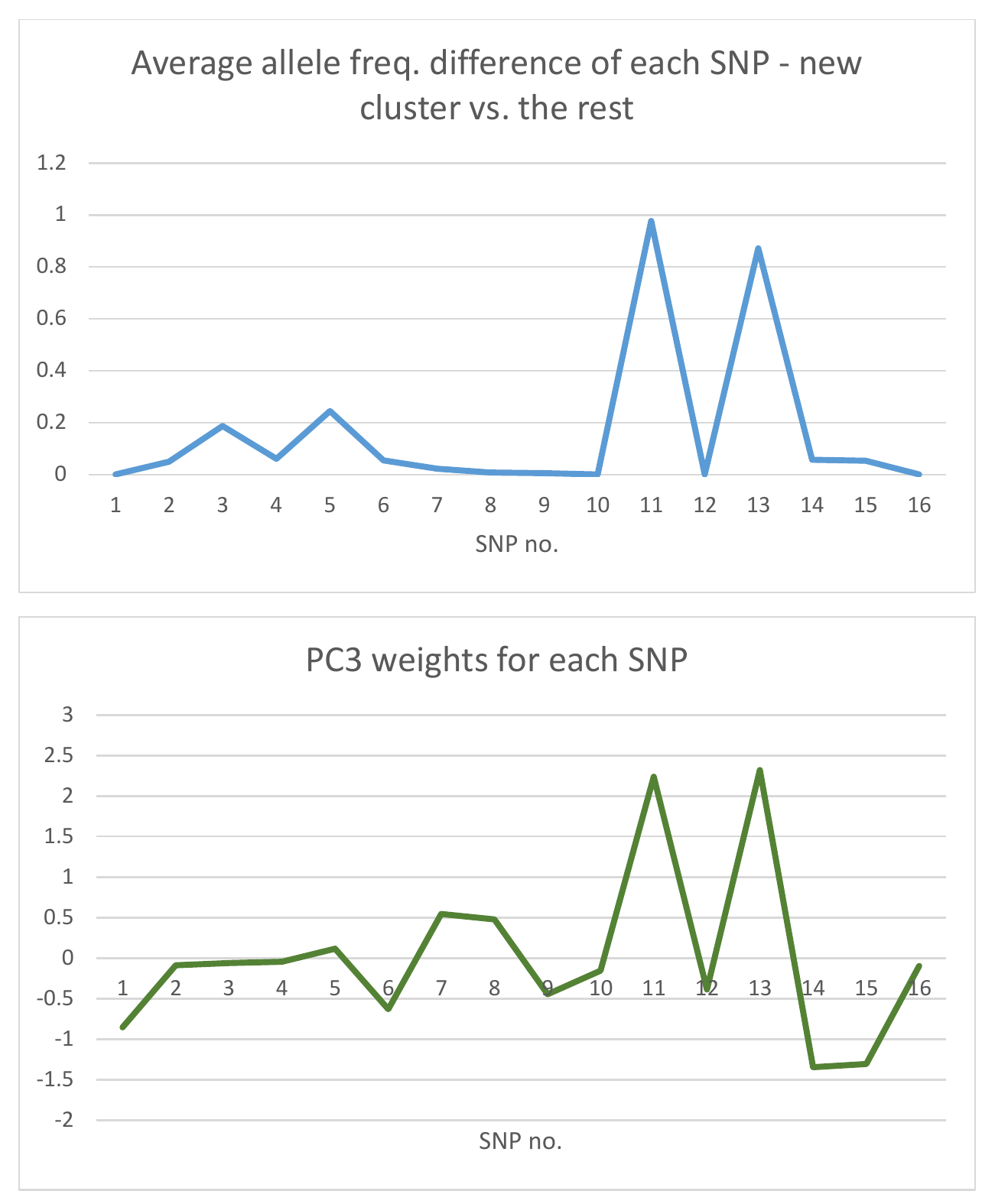} \\
%\begin{tabular}{c}
%\includegraphics[width=0.5\textwidth]{FIGCOLOMBIA/difffreq} \\
%\includegraphics[width=0.5\textwidth]{FIGCOLOMBIA/pcsnp} \\
%\end{tabular}
\caption{Top panel: absolute value of the difference between the average allele frequency for each SNP in all the clusters and the frequency in the rare cluster. Bottom panel: loadings for each SNP of the third PC.}
\end{figure}
\end{center}

\begin{center}
\begin{figure}
\label{fig:adprop}
\includegraphics[width=\textwidth]{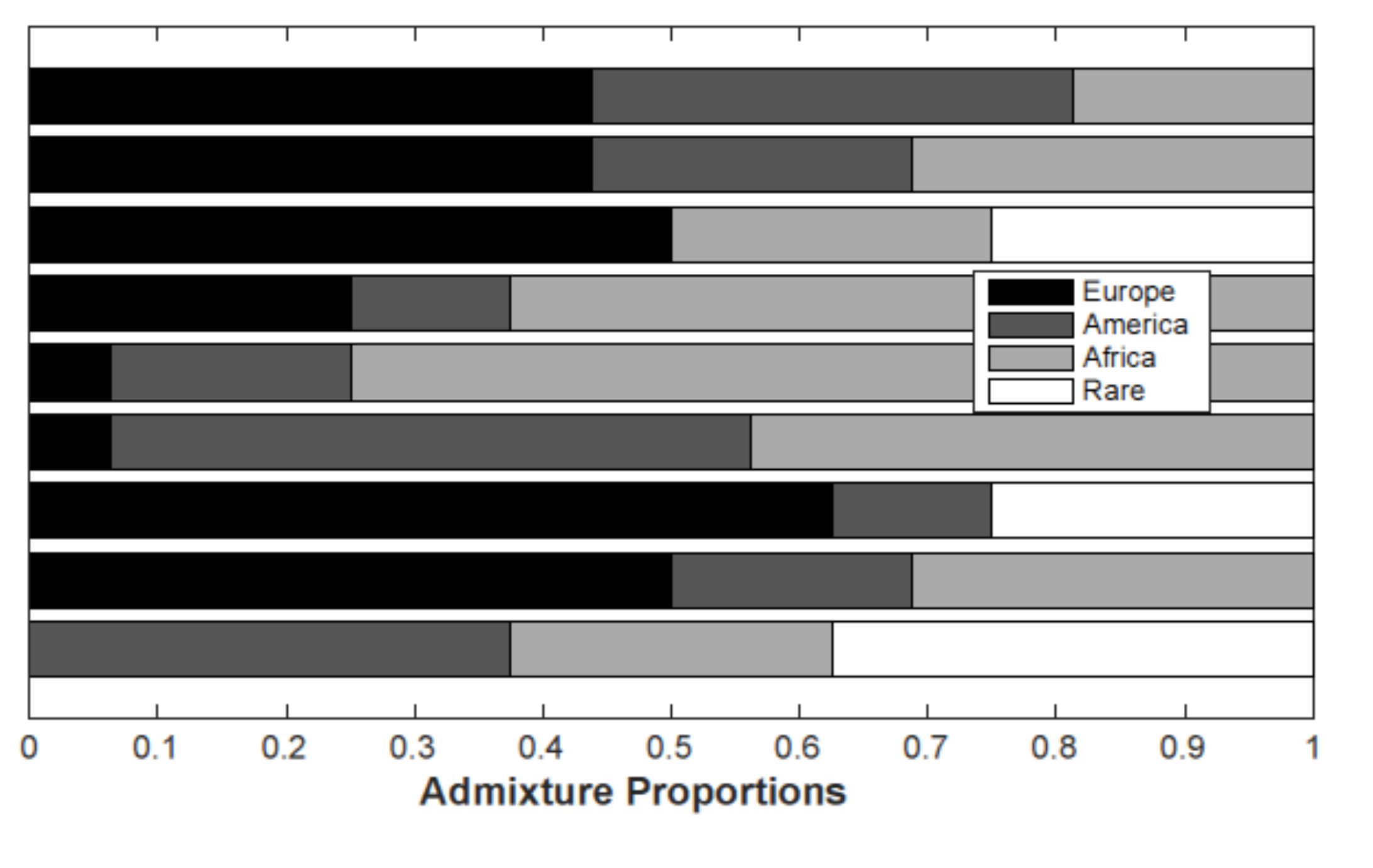}
\caption{Posterior admixture proportions for randomly selected haplotypes.}
\end{figure}
\end{center}

The posterior analysis shows evidence of four major ancestral populations in the set of 744 Colombian haplotypes (see Figure \ref{fig:K}).
We use the MCMC output to estimate the cluster assignment, i.e. population allocation, to each of the 4 major ancestral populations for each haplotype sequence and each marker. In Figure \ref{fig:clus}, we summarise the MCMC output by reporting the clustering that minimizes the posterior expectation of Binder's loss as described by 
Fritsch and Ickstadt (2009)\cite{Fri(09)}, who also discuss possible alternatives such as  
Maximum \emph{a posteriori} clustering. 
The four major clusters have admixture coefficients, i.e. relative proportion of occurrences of each of the clusters, 51.8\%, 32.1\%, 11.4\% and 4.7\% respectively. As we have used a reference panel, we are able to identify in the first cluster, in terms of cardinality, European-origin haplotypes in the sampled Colombians. The second and third clusters correspond to Amerindian and African respectively. This is also confirmed by looking at the ``most frequent" haplotype in each cluster.
%which describe 95\% of the dataset with admixture coefficients 0.50, 0.38, 0.08 and 0.04. The two most common populations can be traced back to Native
%Americans and Europeans; the third can be connected to African ancestry and the presence of a  fourth population  captures  a rare allele in Europeans. This is consistent with what is known about the historical admixture in Colombia.
Figure \ref{fig:dataclus} shows the raw data assigned to each of the four major ancestral populations. We verify our findings in two ways. Firstly, we calculate genetic ancestry proportions using reference genotypes as reference ancestral groups. EDAR-specific ancestry proportions for each of the 372 Colombian samples were estimated using Admixture software \cite{Ale(09)},
which provides a faster implementation of the same model in STRUCTURE. 
We correlated these ancestry proportions to our cluster occurrence proportion (see Table \ref{tab:corr}). The correlation values are very high and support our assignment of ancestry category to the first three clusters. The average European, Amerindian and African ancestry across Colombian samples are 53.6\%, 30.8\% and 16.6\% respectively, which is also very close to our cluster proportions.
\begin{table}
\label{tab:corr}
\begin{tabular}{|c|c|c|c|}
\hline
Correlation	&European anc.	& Amerindian anc.&	African anc.\\ \hline
Cluster 1 (Europe)	&0.90 &	-0.67	&-0.36 \\ \hline
Cluster 2 (America)&	-0.75	&0.92&	-0.20 \\ \hline
Cluster 3 (Africa)	&-0.36	&-0.19&	0.76 \\ \hline
Cluster 4 (New)	& 0.04&	-0.24&	0.28
\\ \hline
\end{tabular}
\caption{Correlation between ancestry proportions and  cluster occurrence proportion from the Bayesian nonparametric model.}
\end{table}
However, Admixture is a supervised approach and so it cannot give us further details about the fourth, rarer cluster. To explore it further, and also for another line of verification, we calculate genetic principal components, in which SNP genotypes for each person is recoded into 0/1/2 by an additive count of the minor allele on two chromosomes, and this SNP genotype count matrix is converted to principal components (PC) via the usual method. As European+Amerindian continental genetic mixing is the primary source for admixture in our data, the first PC axis reflects this, being positively correlated with European samples and negatively with American samples. The second PC captures the other continental component in our samples, namely the African samples. More specifically, PC1 captures the European-Amerindian axis of variation, and PC2 captures the African-European axis. In Table \ref{tab:pc}  we show correlations of PCs with supervised ancestry values. As further PCs are orthogonal to these, they do not show high correlation with any ancestry component.
\begin{table}
\label{tab:pc} 
\begin{tabular}{|c|c|c|c|}
\hline
& European anc. &	Amerindian anc.	& African anc.\\ \hline
PC1 &	0.81	&-0.95	&0.15\\ \hline
PC2	&-0.60	&-0.04	&0.90\\ \hline
PC3	&-0.09	&0.00	& 0.13\\ \hline
PC4	& 0.15	&0.03	&-0.25\\ \hline
\end{tabular}
\caption{Correlations between principal components and  supervised ancestry values.}
\end{table}
Consequently, the first PC shows high correlations with the first two clusters, and PC2 with the third cluster. As shown in Table \ref{tab:pcclus} the third PC is highly correlated with the new cluster, which validates the signal we capture as genuine genetic component and not a statistical artefact of our method.
\begin{table}
\label{tab:pcclus} 
\begin{tabular}{|c|c|c|c|c|}
\hline
Correlation	&PC1	&PC2	&PC3	&PC4 \\ \hline
Cluster 1 (Europe)	&0.73	&-0.58&	-0.26	&0.24\\ \hline
Cluster 2 (America)&	-0.90	&0.04	&0.02&	-0.01\\ \hline
Cluster 3 (Africa)	&0.06&	0.72&	-0.07&	-0.12\\ \hline
Cluster 4 (New)	& 0.24	&0.31&	0.71&	-0.40\\ \hline
\end{tabular}
\caption{Correlations between principal components and  cluster assignment.}
\end{table}
To investigate the genetic source of the new cluster, we look at the average allele frequency for each SNP in all the clusters, and then take the difference for the new cluster vs. all the others. Figure \ref{fig:rareclus}, top panel, shows the absolute differences: we see clearly that only SNPs 11 and 13 primarily contribute to this cluster. The same is seen when we plot the weights given by PC3 onto each SNP (Figure \ref{fig:rareclus} bottom panel). These two SNPs -- rs260693 and rs260696 -- are rare SNPs, i.e. their minor allele which is recognized in cluster 4 and PC3 is rare. For example, rs260693 has a global MAF of 3.8\%, with the minor allele only primarily seen in Europe (9\%) while nearly being absent in Africans (1.8\%) and East Asians (0\%). Their two minor alleles are highly in LD, with a D' of 1 in European populations. This shows that the haplotype that contains the two minor alleles for these two SNPs is also rare, and is being identified as the fourth separate cluster by our model.
Table \ref{tab:maf} reports the posterior mean of the $\theta_{kl}$ of the 16 SNPs in each of the four major populations while  Figure \ref{fig:adprop} shows the posterior mean of admixture proportion $\pi_{ik}$ for few randomly selected individuals.

\begin{table}
\label{tab:maf}
\begin{center}
\includegraphics[width=0.7\textwidth]{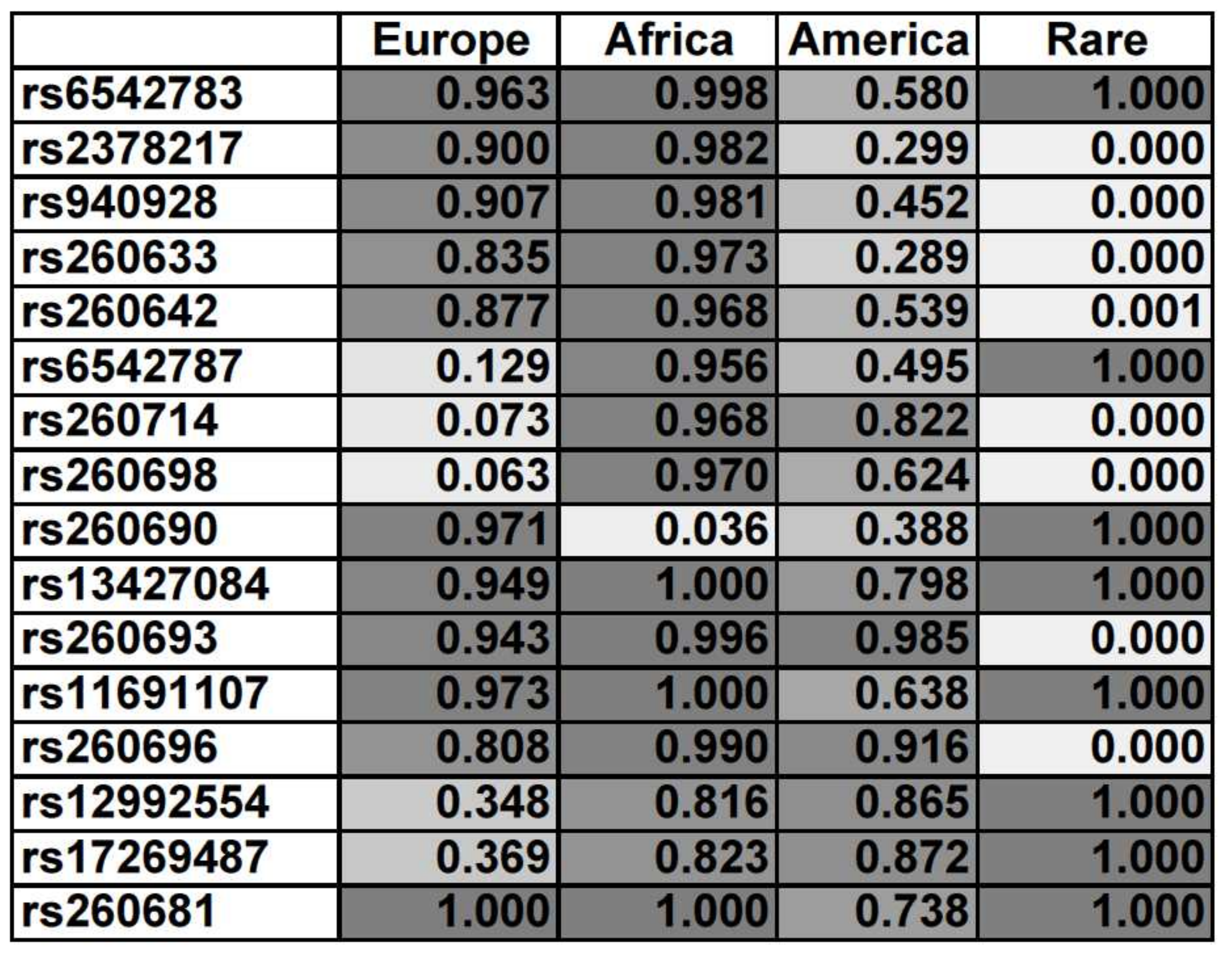}
\caption{Posterior mean of $\theta_{kl},l=1,\ldots,16$  in each of the four major populations. The colour gradient in each cell is proportional to the numerical value.}
\end{center}  
\end{table}

\section{Discussion and concluding remarks}
\label{sec:discussion}

We have presented the Bayesian nonparametric counterpart of the Linkage model of Falush et al.\ (2003)~\cite{Fal(03)} to infer genetic admixtures. The model allows for both the number of ancestral population and the assignment vector to be random, avoiding the use of model selection criteria.  The model can be applied to commonly used genetic markers and does not rely on specific assumption on the mutation model. We incorporate dependence between markers due to correlation of ancestry by specifying an inhomogeneous Poisson process on the DNA sequence.  Each population is modelled using a simple and independent allele-frequency profile. We have developed an MCMC algorithm which allows us to perform posterior inference on
the number of ancestral populations,
the  population of origin of chromosomal regions,
the proportion of an individual's genome coming from
each population,
the admixture proportions in the population and
the allele frequencies in ancestral populations. 
We have demonstrated the model in simulations and on real data from the EDAR gene. The model has been able to highlight the existence of a rare European haplotype.  
We have highlighted possible extensions to our method. An interesting direction for future development is to relax the assumption of independent allele-frequency profile in each population  by incorporating ideas from Sohn et al.\ (2012)~\cite{Soh(12)} and model each population as a hidden Markov model over a set of founder haplotypes. 

In this article we have devoted considerable attention to inferring $K$ and shown how Bayesian nonparamertic methods automatically provide posterior inference on the number of ancestral populations. Nevertheless, we must be careful when interpreting $K$.  The nonparametric setup will generally yield sensible clustering but clusters will not necessarily correspond to ``real" populations.  This problem is in common with most of the model-based structure algorithms \cite{Pri(00)}.

\section*{Appendix}

We develop a Gibbs sampler algorithm, in which each variable is updated given the remaining variables fixed at their current value.
\begin{description}
\item[\textit{Update of $G_0$.}] See section 2.4 of main manuscript.
\item[\textit{Update of $G_i$.}] See section 2.4 of main manuscript.
\item[\textit{Update of $n_{ik}$ and $m_{ik}$.}]  Assume that at iteration $w$ there are $K$ populations in the model. We need to resample the seating arrangement of the Chinese restaurant of $G_i$. Updating $n_{ik}$ is straightforward as the new sample is simply the number of linked segments with $z_{il}=k$. That is,
$$n_{ik}=\sum_{l=1}^L \boldsymbol{1}(s_{il}=0) \boldsymbol{1}(z_{il}=k).$$
Then, $m_{ik}$ can be sampled from the distribution of the number of tables in Chinese Restaurant  Process with $n_ik$ customers and mass parameter
$\alpha q_{0k}$. That is,
\begin{eqnarray*}
m_{ik} &=& \sum_{j=1}^{n_{ik}} b_j\\
b_j & \sim & \Bernoulli\left(\frac{\alpha q_{ik}}{\alpha q_{ik} + j-1}\right)
\end{eqnarray*}
where $b_j=1$ if customer $j$ joins a new table. 
\item[\textit{Update of $\theta_{kl}$.}] 
Assume that at iteration $w$ there $K$ populations in the model. The posterior $p(\theta_{kl}) \mid \text{rest}) $ is the same as in a simple parametric Bayesian model using as observations all the markers for which $z_{il}=k$. In the case of SNP data the conditional posterior of $\theta_{kl}$ is a Beta distribution.   
\item[\textit{Update of $\alpha$.}] The concentration parameter $\alpha_0$ governs the distribution of the number of of $\theta_{kl}$'s in each mixture. We follow Teh et al.\ (2006)~\cite{Teh(06)}.  Assume that at iteration $w$ there $K$ populations in the model. Let $m_{..}=\sum_{i,k} m_{ik}$ 
and $n_{i.}= \sum_k n_{ik}$. We introduce latent variable $w_i \in [0,1]$ and $t_i\in \{0,1\}$, $i=1,\ldots,N$, with
\begin{eqnarray*}
w_i \mid \alpha_0 & \sim & \Bet(\alpha_0 + 1, n_{i.}) \\
t_i \mid \alpha_0 & \propto & \left( \frac{n_{i.}}{\alpha_0+ n_{i.}}\right)^{t_i}. 
\end{eqnarray*}
If $\alpha$ is given a $\Gamma(a,b)$ hyperprior, then given $w_i$ and $t_i$,  $\alpha$ has a Gamma distribution with parameters  $a+m_{..} - \sum_{i =1}^N t_i $ and $b- \sum_{i =1}^N \log w_i$.
\item[\textit{Update of $\alpha_0 $.}] Given the total number $m_{..}=\sum_{i,k} m_{ik}$ of the $\theta_{kl}$'s, the concentration parameter $\alpha_0$ governs the distribution of the number of population $K$. We use the auxiliary method of Escobar and West~\cite{Esc(95)}. If $\alpha_0$ is given a $\Gamma(a,b)$ hyperprior, it can be resampled by introducing a latent variable $\gamma\sim \Bet (\alpha_0+1,m_{..})$ and
$$ p(\alpha_0 \mid K,\gamma) = \pi \Gamma(a+ K, b-\log(\gamma)) +(1-\pi) \Gamma(a+ K-1, b-\log(\gamma)) $$ 
where $\pi/(1-\pi)= (a+K-1)/m_{..}(b-\log(\gamma))$.
\item[\textit{Update of $r $.}] 
The rate $r$ of the Poisson process is given   a Uniform prior on some interval $[r_L,r_U]$. We use a  random walk Metropolis step to update $r$ with proposal distribution centred around the current value. 
\item[\textit{Update of hyperparameters in the base measure $H $.}] The proportion of the model involving the hyperparameters in $H$ is a convential parametric model. Hence, conditioning on all the other variables usually leaves us with a standard Bayesian model, often in conjugate form. In the case of SNP data, we have taken $\theta_k\sim H=\prod_{l=1}^L \Bet(c \mu_l, c(1-\mu_l))$. We specify independent $\Bet(a_l,b_l)$ for each $\mu_l$, $l=1,\ldots,L$, and update $\mu_l$  using a  random walk Metropolis step  with proposal distribution centred around the current value.
\end{description}

\newpage
%%%%%%%%%%%%% FIGURES

\begin{center}
\begin{figure}
\label{fig:structres}
\includegraphics[width=\textwidth]{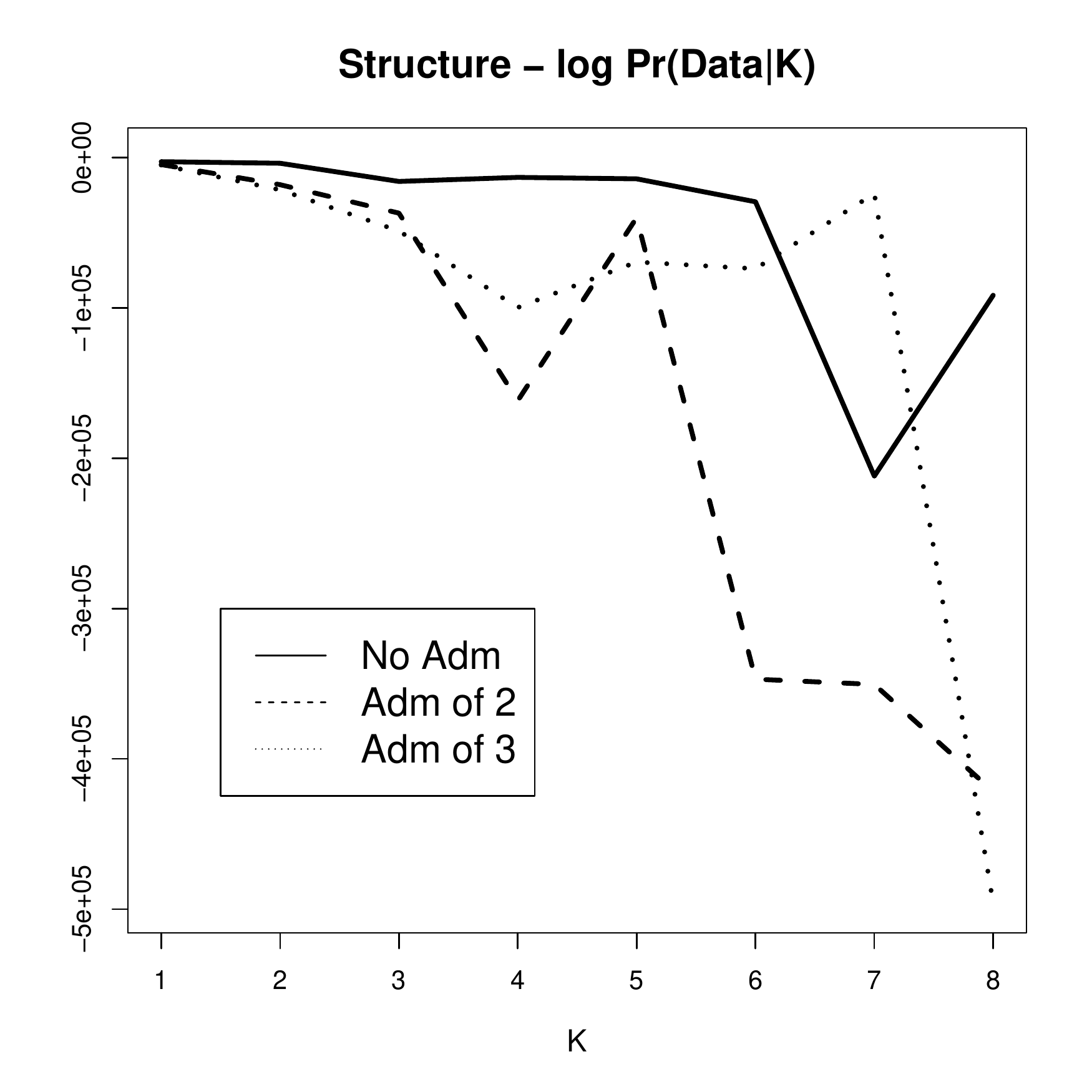}
\caption{STRUCTURE software: $\log \Pr( \text{Data}\mid )$ under the three simulated scenarios.}
\end{figure}
\end{center}

\end{document}